\documentclass[11pt,a4paper]{article}
\usepackage{jcappub}
\usepackage{graphicx}
\usepackage{dcolumn}
\usepackage{bm}

\newcommand{\be}{\begin{equation}}
\newcommand{\en}{\end{equation}}
\newcommand{\bea}{\begin{eqnarray}}
\newcommand{\ena}{\end{eqnarray}}

\title{Cosmic slowing down of acceleration for several dark energy parametrizations}
\author[a]{Juan Maga\~na}%
\emailAdd{juan.magana@uv.cl}
\author[a]{V\'ictor H. C\'ardenas}
 \emailAdd{victor.cardenas@uv.cl}
 \author[a]{Ver\'onica Motta}
 \emailAdd{veronica.motta@uv.cl}

\affiliation[a]{Instituto de F\'{\i}sica y Astronom\'ia, Facultad
de Ciencias, Universidad de Valpara\'iso, Avda. Gran Breta\~na 1111,
Valpara\'iso, Chile.
}%

\date{\today}

\abstract{
We further investigate slowing down of acceleration of the universe scenario
for five parametrizations of the equation of state of dark energy
using four sets of Type Ia supernovae data. In a maximal probability analysis 
we also use the baryon acoustic oscillation and cosmic
microwave background observations. We found the low redshift
transition of the deceleration parameter appears, independently of
the parametrization, using supernovae data alone except for the Union
2.1 sample. This feature disappears once we combine the Type Ia supernovae
data with high redshift data. We conclude that the rapid variation
of the deceleration parameter is independent of the parametrization.
We also found more evidence for a tension among the supernovae
samples, as well as for the low and high redshift data. }

\keywords{dark energy, Cosmology}

\begin{document}
\maketitle

\section{Introduction}
The observations from Type Ia supernova (SNIa) were the first clear
evidence of the acce-lerated expansion of the Universe
\cite{Riess:1998cb,Perlmutter:1998np}. 
The hypothesis of an expanding universe is currently supported by
a wide range of large-scale measurements
\cite{Hinshaw:2012, planckXVI, Davis:2014csa}.
To explain this acceleration within general relativity, it is necessary to introduce an exotic
component, named dark energy (DE). Current observations point
towards a cosmological constant $\Lambda$ \cite{Serra:2009, Nair:2013, Postnikov:2014aua};
nevertheless, it could be necessary to consider dynamical DE models
where the equation of state (EoS) parameter, $w$, is a function of
the scale factor $a$ (or redshift $z$), i.e. $w=P/\rho=w(a)$, where
$P$ and $\rho$ are the pressure and energy density of DE
respectively \cite{Wetterich:1987fm,weinberg89,cop06, Vazquez:2012ce}. One popular ansatz for this
dynamical EoS is the Chevallier-Polarski-Linder (CPL)
parametrization \cite{Chevallier:2000qy,Linder:2003nc} given by
$w(a)=w_{0}+(1-a)w_{1}$, where $w_{0}$ is the present value of the
EoS and $w_{1}$ is the derivative with respect to scale factor.
Recently, \cite{Shafieloo:2009} found a cosmic slowing down of the
acceleration at low redshifts using the CPL model with the
Constitution SNIa sample in combination with baryon acoustic
oscillation (BAO) data. Nevertheless, this apparent behavior
disappears when the cosmic microwave background (CMB) data are added
in the analysis, indicating a tension between low and high redshift
measurements. Similarly, using the Union2 SNIa set, \cite{Li:2011} it was
found the same behavior at low redshift. The authors of
\cite{victor_rivera} found the tension between the sets at low
redshifts and at high redshifts can be ameliorated incorporating a
curvature term as a free parameter in the analysis.

The slowing down scenario using only supernovas can be considered a
sort of an artifact from the data. However, recent studies have been
able to demonstrate the same trend using using $42$ measurements of
gas mass fraction in galaxy clusters \cite{victor_fgas}. They found the same trend
as using SNIa, i.e. the acceleration of the universe
has already reached its maximum at $z\sim 0.2$ and is currently
moving towards a decelerating phase, in agreement with previous
works. At this point it would be interesting to know if this feature
is only an artifact of the CPL parametrization. 

In this paper we study the cosmic slowing down of the acceleration
at low redshifts using four SNIa sets: Constitution, Union 2, Union
2.1, and LOSS-Union, together with five parametrizations of the EoS
of dark energy. We also consider the constraints from BAO and CMB
from the Wilkinson Microwave Anisotropy Probe (WMAP) 9-yr
observations.

The paper is organized as follows: in the next section we introduce
the cosmological framework for a flat universe case. In section \S
\ref{sec:par} we present the dynamical models of dark energy. We
describe in \S \ref{sec:data} the SNIa data used to constrain the
parameters of the models and present our results. Finally, we
discuss the results and present our conclusions in section \S
\ref{sec:conclusions}

\section{The cosmological framework} \label{sec:method}
We consider a flat Friedmann-Lema\^itre-Robertson-Walker (FLRW)
universe whose DE has a dynamical EoS $w(z)$. The dimensionless
Hubble parameter $E(z)$ for this universe is given by
\begin{equation}
 E^{2}(z)=H^{2}(z)/H^{2}_{0}=\Omega_{m}(1+z)^{3} + \Omega_{r}(1+z)^{4} +\Omega_{de}X(z),
 \label{eq:Ez}
\end{equation}
where $\Omega_{m}$ is the density parameter for matter. The density
parameter for radiation is
$\Omega_{r}=2.469\times10^{-5}h^{-2}(1+0.2271 N_{eff})$, where
$h=H_{0}/100\, \mathrm{kms}^{-1}\mathrm{Mpc}^{-1}$ and the number of
relativistic species is set to $N_{eff}=3.04$ \cite{Komatsu:2010fb}. The density parameter
for DE is written as $\Omega_{de}=1-\Omega_{m}-\Omega_{r}$ and the
function $X(z)$ reads as
\begin{equation}
X(z)=\frac{\rho_{de}(z)}{\rho_{de}(0)}=
\mathrm{exp}\left(3\int^{z}_{0}\frac{1+w(z)}{1+z}\mathrm{dz}\right),
\label{eq:fz}
\end{equation}
where $\rho_{de}(z)$ is the energy density of DE at redshift $z$, and
$\rho_{de}(0)$ its present value. The comoving distance from the
observer to redshift $z$ is given by
\begin{equation}
r(z)=\frac{c}{H_0}\int_0^z \frac{dz'}{E(z')}.
\label{eq:rz}
\end{equation}
%
%
The deceleration parameter \textit{q(z)} is defined as
\begin{equation}
q(z) = - \frac{\ddot{a}(z)a(z)}{\dot{a}^{2}(z)},
\label{eq:qa}
\end{equation}
where $a$ is the scale factor of the Universe. Using eq.
(\ref{eq:Ez}), this expression can be rewritten as
\begin{equation}
q(z) = -\frac{(1+z)}{E(z)} \frac{dE(z)}{dz}-1.
\label{eq:qz}
\end{equation}

In the following section, we study five different parametrizations of $w(z)$ 
and for each one we shall use expression (\ref{eq:qz}) to reconstruct the deceleration parameter.

\section{Parametrizations of the dark energy equation of state} \label{sec:par}

Because we do not known what physics is behind the DE, one way to explore models 
going beyond the cosmological constant is by using an explicit parametrization for 
the EoS parameter $w(z)$. Actually, the apparent cosmic slowing down of acceleration 
was found first using an explicit parametrization of $w(z)$. 
This low redshift feature of the reconstructed $q(z)$, remains after using different data sets 
by applying the CPL parametrization \cite{Shafieloo:2009, victor_fgas}. 
In the \cite{Cardenas:2014jya}, the author adopted a different approach focusing his study not in a
$w(z)$ trial function, but in a simple (blind analysis) interpolation for the DE density $X(z)$. 
However, he found the same low redshift transition for the reconstructed $q(z)$. 
This means that both are valid complementary methods which enable us to explore evolution 
in the DE density, and consequently, the cosmic acceleration.

In this work, although we do not expect that a phenomenological specific parametrization for $w(z)$
will reproduce the eventual real behavior of $q(z)$, we investigate through several 
parametrizations of the DE EoS whether the cosmic slowing down of acceleration is a real feature 
of the Universe at low redshifts. 
We choose parametrizations with two independent free parameters, 
thus allowing for direct statistical model comparisons. 
In the following we introduce the parametrizations to be tested by cosmological data.

\subsection{Jassal-Bagla-Padmanabhan parametrization}

Jassal-Bagla-Padmanabhan (JBP) \cite{Jassal:2004ej,Jassal:2005qc}
propose an EoS of DE as function of redshift in the following way
\begin{equation}
w(z)=w_{0} + w_{1}\frac{z}{\left(1+z\right)^{2}},
\label{eq:JBP}
\end{equation}
where $w_{0}=w(0)$, $w_{1}=w'(0)$, and $w(\infty)=w_{0}$.
In addition, the EoS parameter has the
same value at present and early epochs and performs a rapid
variation around $z=1$ with amplitude $w_0+w_1/4$.
For the JBP parametrization (\ref{eq:JBP}), the function
(\ref{eq:fz}) reads as
\begin{equation}
 X(z)=(1+z)^{3(1+w_{0})}\mathrm{exp}\left(\frac{3}{2}\frac{w_{1}z^{2}}{(1+z)^{2}}\right),
 \label{eq:xjbp}
\end{equation}
and $E(z)$ is completely determined.

\subsection{Feng-Shen-Li-Li parametrization}

Another interesting parametrization was proposed by Feng, Shen, Li
and Li (FSLL) \cite{Feng:2012gf} to overcome the divergence of the
CPL model when $z\rightarrow-1$. They propose the two following relations:
\begin{eqnarray}
 \mathrm{FSLL\;I:}\quad w(z)&=&w_{0} + w_{1}\frac{z}{1+z^{2}},\nonumber\\
 \mathrm{FSLL\;II:}\quad w(z)&=&w_{0} + w_{1}\frac{z^{2}}{1+z^{2}},
 \label{eq:FSLL}
\end{eqnarray}
where for both models $w_{0}=w(0)$ and $w_{1}=w'(0)$. For FSLL I, $w(\infty)=w_{0}$
and at low redshift it reduces to the linear form $w(z)\approx
w_{0}+w_{1}z$, while for FSLL II $w(\infty)=w_{0}+w_{1}$ and at low
redshifts it yields $w(z)\approx w_{0}+w_{1}z^{2}$. For the
parametrizations (\ref{eq:FSLL}), the functions (\ref{eq:fz}) are
given by
\begin{equation}
X(z)_{\pm}=(1+z)^{3(1+w_{0})}\mathrm{exp}\left[\pm \frac{3w_{1}}{2}\mathrm{arctan(z)}\right]
\left(1+z^{2}\right)^{3w_{1}/4}\left(1+z\right)^{\mp 3w_{1}/2},
\label{eq:xfsll}
\end{equation}
where $X_{+}$ and $X_{-}$ correspond to FSLL I and FSLL II
respectively.

\subsection{Polynomial parametrization}

A generalization of the CPL model consists in an expansion in
powers of $(1+z)$ \cite{Weller:2001gf}. One of these polynomial
parametrizations was proposed by Sendra and Lazkoz (SL)
\cite{Sendra:2011pt}
\begin{equation}
w(z)=-1 + c_{1}\left(\frac{1+2z}{1+z}\right) + c_{2}\left(\frac{1+2z}{1+z}\right)^{2},
\label{eq:polynomial}
\end{equation}
where $c_{1}=(16w_{0}-9w_{0.5}+7)/4$, $c_{2}=-3w_{0}+
(9w_{0.5}-3)/4$, $w_{0}$ and $w_{0.5}$ are the values of the
equation of state at $z=0$ and $z=0.5$ respectively. This
description of the EoS in terms of $w_{0.5}$ improved the situation
of the CPL model, which suffers from a significant correlation
between the parameters $w_{0}$ and $w_{1}$ \cite{Wang:2008zh}. The
proposed function is:
\begin{equation}
X(z)=(1+z)^{3(1-8w_{0}+9w_{1})/2}
\mathrm{exp}\left[\frac{3z(w_{0}(52z+40)-9w_{1}(5z+4)+7z+4)}{8(1+z)^2}\right].
\label{eq:xsl}
\end{equation}
Notice the $\Lambda$CDM model is recovered for $w_{0}=w_{0.5}=-1$.

\subsection{Barbosa-Alcaniz parametrization}

We also consider the Barbosa-Alcaniz parametrization
\cite{Barboza:2008rh}, where:
\begin{equation}\label{ba}
w(z)=w_0 + w_1 \frac{z(1+z)}{1+z^2},
\end{equation}
where $w_{0}=w(0)$, $w_{1}=w'(0)$, $w(\infty)=w_{0}+w_{1}$, and
at low redshift it reduces to the linear form $w(z)\approx w_{0}+w_{1}z$.
This ansatz is a well behaved function of the redshift $z$ over the range
$z\in [-1, \infty)$. The DE density in this case evolves as
\begin{equation}\label{deba}
X(z)=(1+z)^{3(1+w_0)}(1+z^2)^{3w_1/2}.
\end{equation}

As a closing for this section, we have to mention that an important
condition that a given parametrization must fulfill, is that the DE
density has to be lower that the matter energy density in the past.
This is easily checked using the explicit expressions we have
derived, finding that all of our best fit functions satisfies this
requirement.

\section{Cosmological constraints from observational data} \label{sec:data}

The maximal probability analysis we perform to constrain the parameters 
of the different parametrizations considers cosmological observations at different redshifts:
Type Ia supernovae data, the acoustic peaks of BAO and the WMAP 9-yr distance posterior.
In all our analysis, we assume a flat geometry, and a fixed reduced Hubble constant $h=0.697$
\citep{Hinshaw:2012}, thus leaving $\Omega_{m}$, $w_{o}$, and $w_{1}$
($w_{0.5}$ for the SL model) as the only free parameters of the problem.
For the supernova data we use the Constitution (C) set consisting of $397$ SNIa points
\cite{Hicken:2009dk} covering a redshift range $0.015<z<1.551$,
Union$~2$ (U2) consisting in 557 points in the redshift range
$0.511<z<1.12$ \cite{Amanullah:2010vv}, and Union$~2.1$ (U21) which
consists of $580$ points in the redshift range $0.015<z<1.41$
\cite{Susuki:2012}. We also use the sample presented by
\cite{Ganeshalingam:2013mia} consisting in $586$ SNIa in the
redshift range $0.01-1.4$ which considers $91$ points of the Lick
Observatory Supernova Search (LOSS) sample
\cite{Ganeshalingam:2010}. Hereafter, we refer to it as LOSS-Union
(LU) set.

The SNIa samples give the distance modulus as a function of redshift
$\mu_{obs}(z)$ and its error $\sigma_{\mu}$. Theoretically, the distance modulus is computed as
\begin{equation}
\mu(z)=5\log_{10}[d_L(z)/\texttt{Mpc}]+25,
\label{eq:mu}
\end{equation}
which is a function of the cosmology through the luminosity distance
(measured in Mpc)
\begin{equation}\label{dlzf}
d_L(z)=(1+z)r(z),
\end{equation}
valid for a flat universe with $r(z)$ given by Eq. (\ref{eq:rz}). We
fit the SNIa with the cosmological model by minimizing the $\chi^2$
value defined as
\begin{equation}
\chi^2_{SNIa}=\sum_{i=1}^{N}\frac{[\mu(z_i)-\mu_{obs}(z_i)]^2}{\sigma_{\mu_{i}}^2}.
\end{equation}
We also consider the data from BAO and CMB. 
The BAO measurements considered in our analysis are the following:
three acoustic parameter measurements at $z=0.44$, $0.6$, and $0.73$ from the WiggleZ
experiment\citep{Blake:2011en}, two BAO distance measurements
at $z=0.2$, and $0.35$ from the Sloan Digital Sky Survey Data Release 7 (SDSS DR7) \citep{Percival:2010}, 
and one data point at $z=0.106$ from the Six-degree-Field Galaxy Survey
(6dFGS) \citep{Beutler:2011hx}. Thus, the total $\chi^{2}$ for all the BAO data sets can be written
as $\chi^{2}_{BAO}=\chi^{2}_{WiggleZ}+\chi^{2}_{SDSS}+\chi^{2}_{6dFGS}$ (see \cite{victor_fgas} 
for more details to construct $\chi^{2}_{BAO}$). The CMB information considered is
derived from the $9$ years WMAP data \cite{Hinshaw:2012}. 
In our analysis we compare the measurements of the acoustic scale, the shift parameter
and the redshift of decoupling with the theoretical ones
(see \cite{victor_fgas} for details to construct $\chi^{2}_{CMB}$). 
We constrain the cosmological parameters for two cases: using only the Type Ia supernovae
data, where the $\chi^{2}_{Tot}=\chi^{2}_{SNIa}$; and 
using all data (SNIa+BAO+CMB ), where $\chi^{2}_{Tot}=\chi^{2}_{SNIa}+\chi^{2}_{BAO}+\chi^{2}_{CMB}$.
We minimize these $\chi^{2}_{Tot}$ functions with respect to the parameters $\{\Omega_{m}, w_{0}, w_{1}(w_{0.5})\}$
to compute the best estimated values and their errors. 

\begin{figure}[ht]
\noindent\makebox[\textwidth]{%
  \begin{tabular}{cc}
    \includegraphics[width=0.48\textwidth]{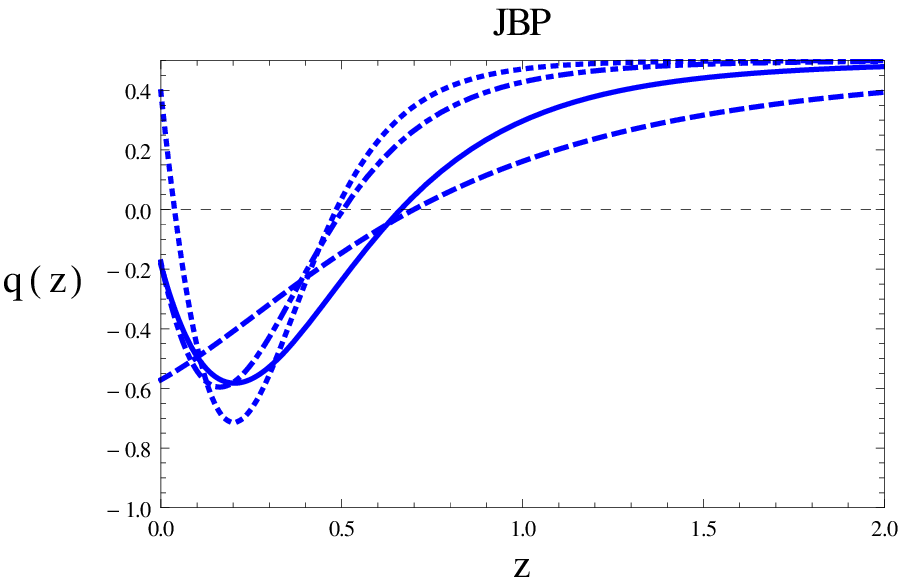} & \includegraphics[width=0.48\textwidth]{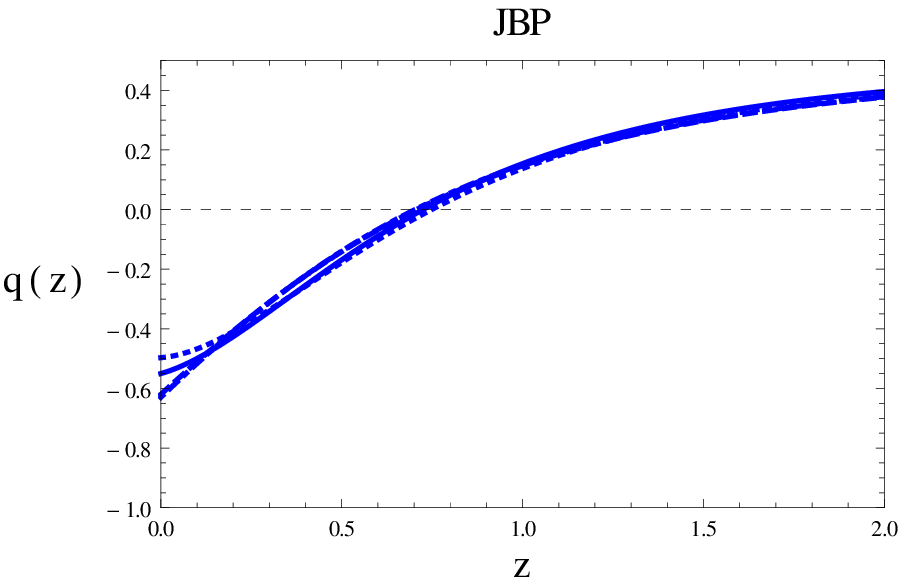}
  \end{tabular}
 }
  \caption{Evolution of the deceleration parameter $q(z)$ vs $z$ for the JBP parametrization. The left panel shows the reconstructed
  $q(z)$ using the best fit obtained from Constitution (dotted line), Union$~2$ (dot-dashed line),
  Union$~2.1$ (dashed line) and LOSS-Union (solid line) samples. The right panel shows the reconstructed
  $q(z)$ using the best fit obtained from C+BAO+CMB (dotted line), U2+BAO+CMB (dotted-dash line),
  U21+BAO+CMB (dashed line) and LU+BAO+CMB (solid line). Notice the cosmic slowing down of the acceleration
  at $z\sim$0.2 for the SNIa sets excepts for Union2.1 sample. This feature disappears when adding BAO and CMB data.}
\label{fig:JBP}
\end{figure}

Our results from the Bayesian analysis for the models are given in Table \ref{tab:table01}.
In Fig. \ref{fig:JBP} we show the evolution of the deceleration parameter as
function of redshift for the JBP parametrization using the best fit values for these data sets.
Notice that using the C set the Universe pass through a maximum of
acceleration at $z\sim 0.2$ and now evolves towards a decelerating
phase in the near future. The same slowing down of acceleration
occurs using U2 (at $z\sim 0.17$), and LU (at $z\sim 0.2$) sets.
This feature is preserved when propagating the error at $1\sigma$
in the best fit parameter. Therefore, there is a statistically significant 
evidence of the cosmic slowing down of the acceleration at low redshifts. 
Nevertheless this behavior does not occur using the U2.1 data. 
When adding the information from BAO and CMB data, the Universe has a
transition from a decelerated phase to an accelerated phase at $z\sim
0.7$ (for all SNIa samples). Our analysis strongly suggests a
tension between low-redshifts and high-redshifts measurements. This
results is in agreement with the previous works using SNIa data and
information from galaxy clusters
\cite{Shafieloo:2009,Li:2011,victor_rivera,victor_fgas}.

\begin{figure}[ht]
\noindent\makebox[\textwidth]{%
  \begin{tabular}{cc}
    \includegraphics[width=0.48\textwidth]{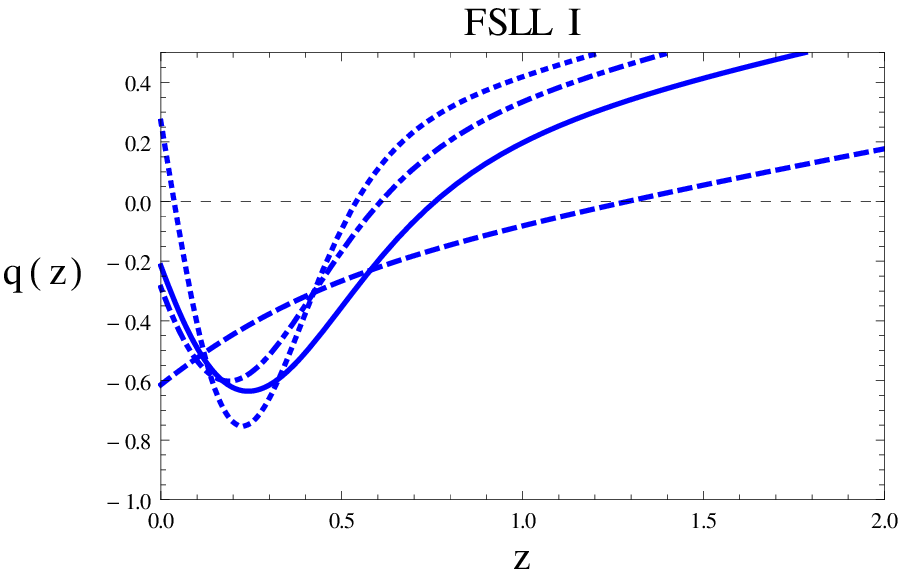} & \includegraphics[width=0.48\textwidth]{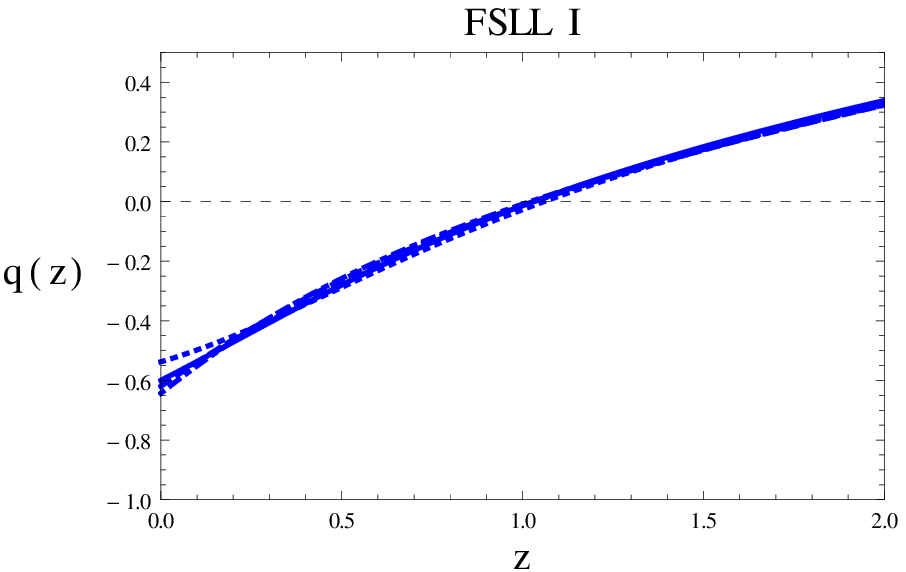}
  \end{tabular}
 }
  \caption{We show the slope of the $q(z)$ as in Fig. \ref{fig:JBP} for the FSLL I parametrization.}
  \label{fig:fsllI}
\end{figure}

\begin{figure}[t]
\noindent\makebox[\textwidth]{%
  \begin{tabular}{cc}
    \includegraphics[width=0.48\textwidth]{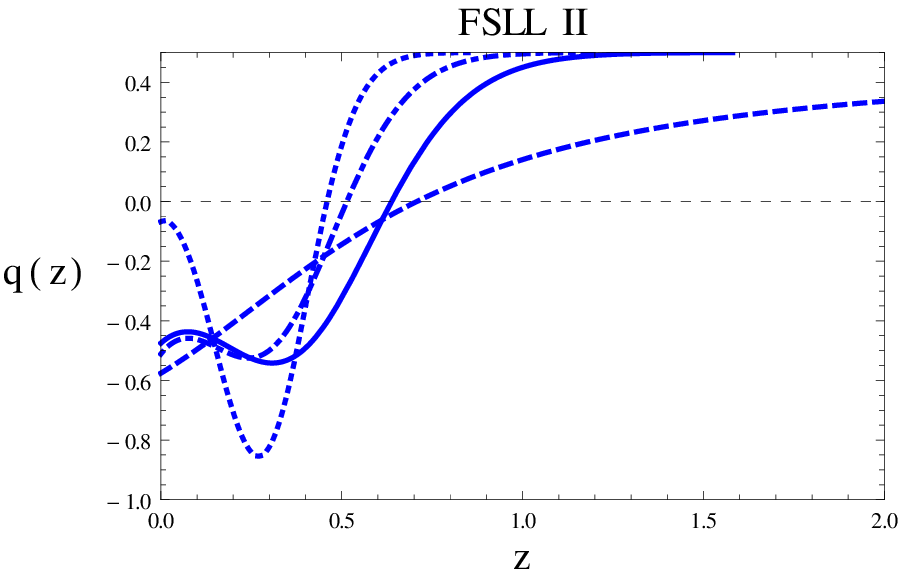} & \includegraphics[width=0.48\textwidth]{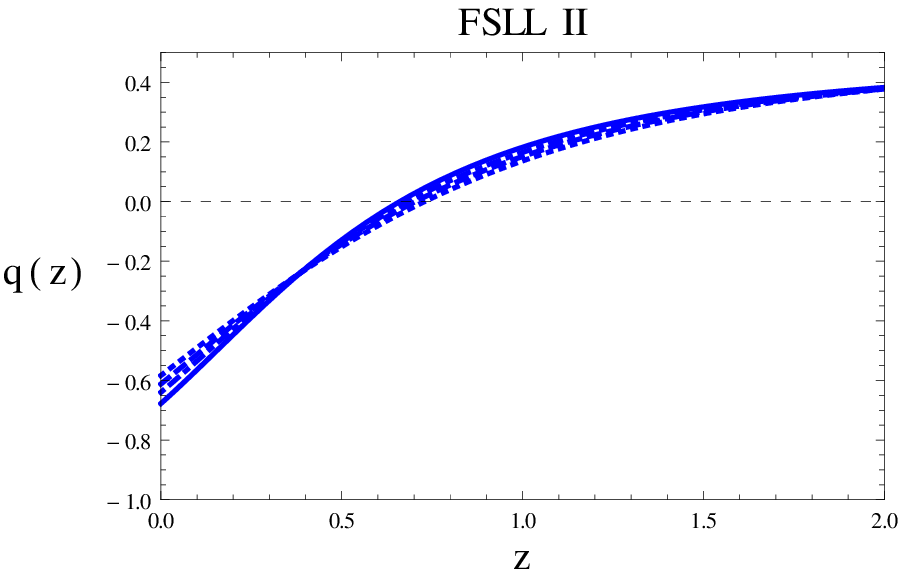}
  \end{tabular}
 }
  \caption{We show the slope of the $q(z)$ as in Fig. \ref{fig:JBP} for the FSLL II parametrization.}
  \label{fig:fsllII}
\end{figure}

\begin{figure}[t]
\noindent\makebox[\textwidth]{%
  \begin{tabular}{cc}
    \includegraphics[width=0.48\textwidth]{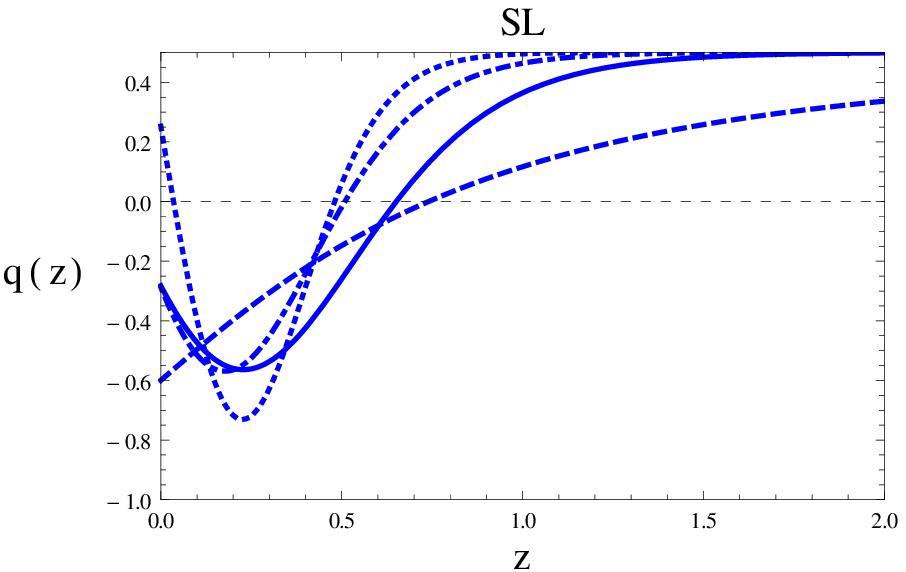} & \includegraphics[width=0.48\textwidth]{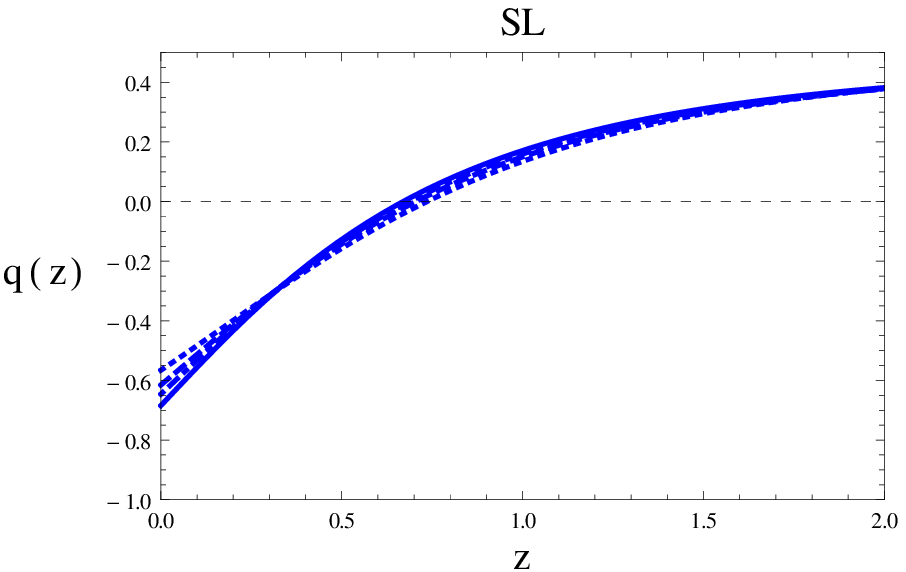}
  \end{tabular}
 }
  \caption{Idem as Fig. \ref{fig:JBP} for the SL parametrization.}
  \label{fig:sl}
\end{figure}

\begin{figure}[t]
\noindent\makebox[\textwidth]{%
  \begin{tabular}{cc}
    \includegraphics[width=0.48\textwidth]{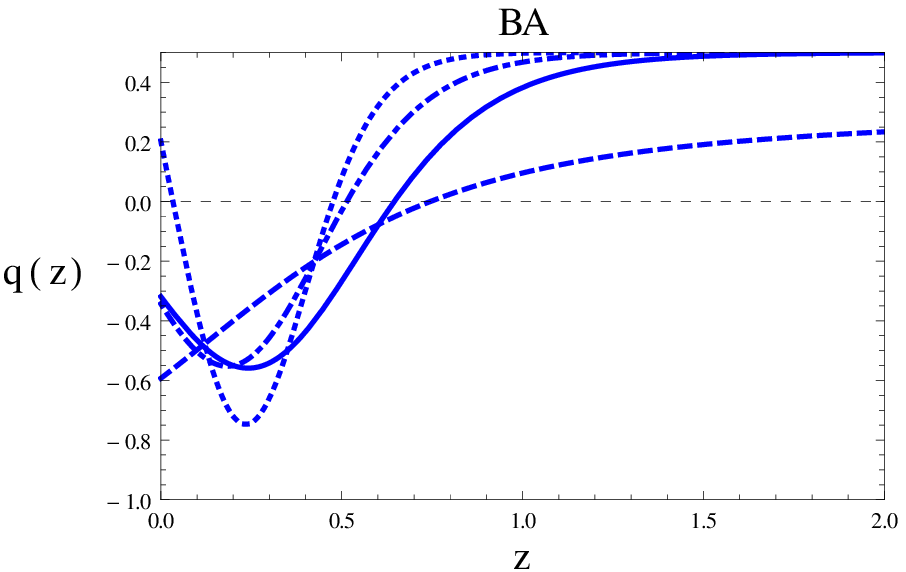} & \includegraphics[width=0.48\textwidth]{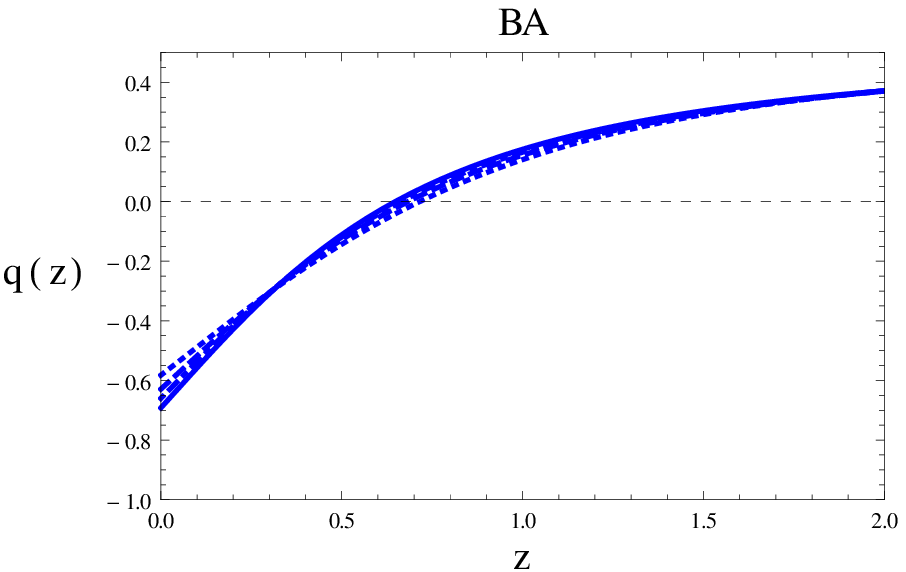}
  \end{tabular}
 }
  \caption{We show the slope of the $q(z)$ as in Fig. \ref{fig:JBP} for the BA parametrization.}
  \label{fig:ba}
\end{figure}

In Fig. \ref{fig:fsllI} we show the reconstructed $q(z)$ as function of $z$ for the FSLL I model.
Notice that the slowing down of the acceleration
occurs at $z\sim0.22$, $z\sim 0.18$, $z\sim 0.25$ using C, U2 an LU
samples respectively. For the FSLL II parametrization, the
cosmic deceleration occurs at $z\sim 0.27$,
$z\sim 0.25$, $z\sim 0.3$ using the same SNIa sets respectively (Figure \ref{fig:fsllII}).
Nevertheless $q(z)$ has another transition at $z\sim 0.07$ for U2 and
LU sets, and the Universe begins an accelerated phase.
However, this oscillating behavior is not statistically significant when
propagating the error at $1 \sigma$ in the best fit parameters.
Using the U2.1 SNIa sample there is no evidence for the deceleration of the Universe at low redshift.
As is shown in Figure \ref{fig:sl} the slowing
down of the acceleration for the SL $w(z)$ function occurs at $z\sim0.22$, $z\sim 0.18$, $z\sim
0.23$ using C, U2 and LU samples respectively.
Finally, in Fig. \ref{fig:ba} we show that the Universe has a maximum of acceleration and it begins to decelerate at
$z\sim0.24$, $z\sim 0.19$, $z\sim 0.25$ using the same SNIa samples respectively.

Therefore, our analysis shows significant evidence for a cosmic slowing down of
the acceleration at low redshifts for dynamical DE models with EoS
varying with redshift using Constitution, Union 2 and LOSS-Union
SNIa sets. Interestingly, this feature disappears when using the
Union 2.1 sample, which suggests an important tension between
this particular data set and the other SNIa samples.
Moreover, when the analysis is performed using the combination of
SNIa data with the BAO and CMB observations, the scenario of the
deceleration of the Universe's expansion at low redshift disappears.
Actually, we obtain $w(0)\approx-1$ at $1\sigma$ for all parametrizations in
agreement with a cosmological constant. Moreover, we found that the Universe has a
transition from a decelerated phase to an accelerated phase at $z\sim1$, $z\sim0.7$,
$z\sim0.75$ and $z\sim0.7$ for the FSLL I, FSLL II, SL, and BA parametrizations respectively.
Therefore, we found more evidence pointing out a tension between the low and high redshift
data.

\begin{figure}[ht]
\noindent\makebox[\textwidth]{%
  \begin{tabular}{cc}
    \includegraphics[width=0.48\textwidth]{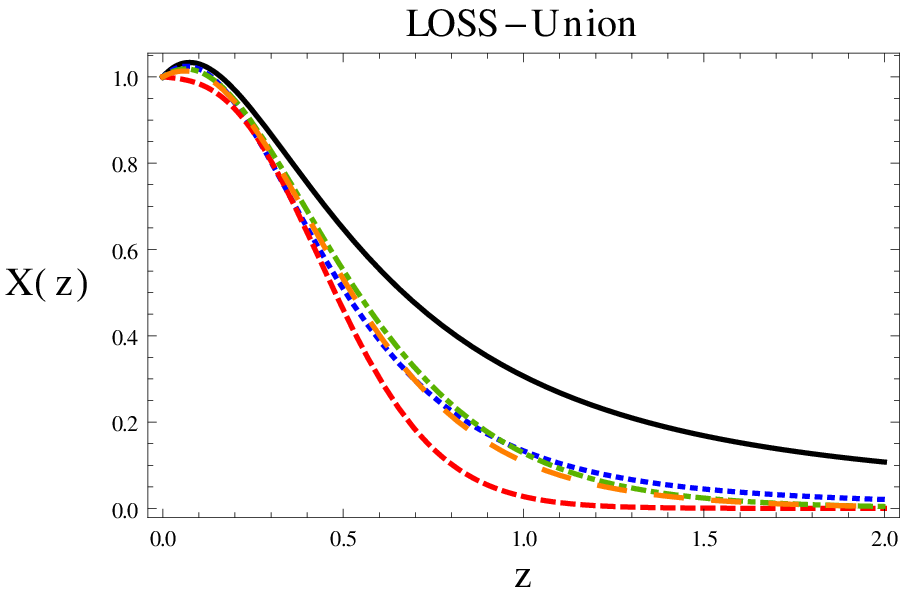} & \includegraphics[width=0.48\textwidth]{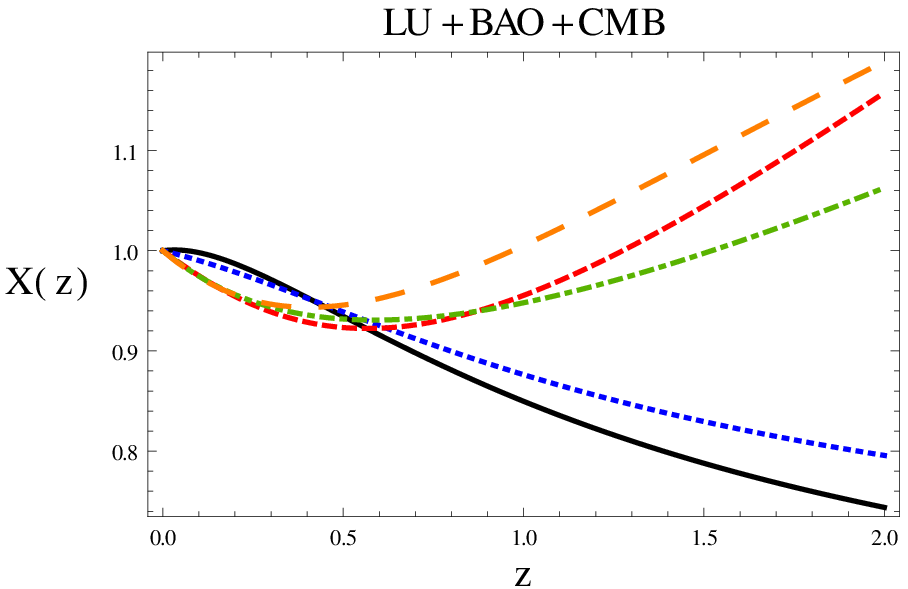}
  \end{tabular}
 }
  \caption{Evolution of the function $X(z)$ vs $z$ for all parametrizations. 
  The left panel shows the reconstructed $X(z)$ using the best fit obtained from LOSS-Union
  for the JBP (solid line), FSLL I (dotted line), FSLL II (dashed), SL (dot-dashed) and,
  BA (long dashed) parametrizations. The right panel shows the reconstructed
  $X(z)$ using the best fit obtained from LOSS-Union+BAO+CMB for the five parametrizations.  
  Notice the low redshift data suggest the DE density is in fact increasing with time.
  This same behavior occurs using the constraints derived from the other Type Ia supernovae data.}
\label{fig:Xz}
\end{figure}

Our results are in agreement with previous work
\cite{Shafieloo:2009, victor_rivera, victor_fgas}, showing that even
with more new data points, the tension between the low
and hight redshift observational probes remains.  One of the main
result of this work is to remove the uncertainty about the effect of
the parametrization of $w(z)$ on the results and final conclusions.
Recently \cite{Cardenas:2014jya} found a
connection between this low redshift transition of $q(z)$
with the DE density decreasing with increasing redshift.
Therefore, it would be important to check such a behavior. 
In Figure \ref{fig:Xz} we show the slope of the $X(z)$ function 
for all the five parametrizations using the best fit parameters estimated from the LOSS-Union
sample. The low redshift data suggest the DE density is in fact increasing with time. 
This same behavior occurs using the constraints derived from the other Type Ia supernovae data.
As it was mentioned in \cite{Cardenas:2014jya} this particular results is in agreement
with the recent BAO DR11 measurements \cite{Delubac:2014aqe} where a
tension between BAO data and CMB was informed. This tension reveals
that, in order to accommodate these new data assuming a flat
universe with dark matter and DE, we have to assume the DE density
decreases with increasing redshift, and it reaches a negative value
at $z\simeq 2.4$,
\begin{equation}
\frac{\rho_{de}(z=2.34)}{\rho_{de}(z=0)}=-1.2 \pm 0.8,
\end{equation}
in agreement with what we have found in this work. Similar
conclusions are obtained in \cite{Sahni:2014ooa} based on the same
data from BAO.


\begin{table}
\centering
\begin{tabular}{lcccc}
\hline
Data Set&$\chi^2_{min}/d.o.f.$ &$\Omega_{m}$ & $w_{0}$ & $w_{1} (w_{0.5})$ \\
\hline
\multicolumn{5}{c}{JBP parametrization}\\
\hline
Constitution&$461.597/394$ & $0.444 \pm 0.048$& $-0.125 \pm 0.647$ &$ -13.820 \pm 8.510$\\
C+BAO+CMB&$467.778/403$ & $0.284 \pm 0.008 $& $-0.929 \pm 0.190$ & $-0.649 \pm 1.383$\\
Union 2&$541.338/554$  & $0.414 \pm 0.072$& $-0.769 \pm 0.439$ &$ -7.201 \pm 7.659$\\
U2+BAO+CMB&$544.752/563$  & $0.283 \pm 0.008$& $-1.051 \pm 0.164$ &$ 0.144 \pm 1.173$\\
Union 2.1&$562.222/577$ & $0.294 \pm 0.262$& $-1.011 \pm 0.199$ &$ -0.274 \pm 5.717$\\
U21+BAO+CMB&$564.364/586$ & $0.283 \pm 0.008$& $-1.041 \pm 0.156$ &$ 0.118 \pm 1.108$\\
LOSS-Union&$573.643/583$ & $0.333 \pm 0.099$& $-0.684 \pm 0.289$ &$ -4.908 \pm 5.503$\\
LU+BAO+CMB&$581.007/592$ & $0.288 \pm 0.009$& $-0.984 \pm 0.162$ &$ -0.523 \pm 1.275$\\
\hline
\multicolumn{5}{c}{FSLLI parametrization}\\
\hline
Constitution&$461.099/394$ & $0.452 \pm 0.042$& $-0.280 \pm 0.546$ &$ -9.472 \pm 5.503$\\
C+BAO+CMB&$467.870/403$ & $0.284 \pm 0.008 $& $-0.966 \pm 0.138$ & $-0.235 \pm 0.616$\\
Union 2&$541.476/554$  & $0.419 \pm 0.069$& $-0.904 \pm 0.348$ &$ -4.510 \pm 4.989$\\
U2+BAO+CMB&$544.697/563$  & $0.283 \pm 0.008$& $-1.063 \pm 0.124$ &$ 0.141 \pm 0.536$\\
Union 2.1&$562.208/577$ & $0.230 \pm 0.610$& $-0.965 \pm 0.630$ &$ 0.388 \pm 3.294$\\
U21+BAO+CMB&$564.356/586$ & $0.282 \pm 0.008$& $-1.041 \pm 0.119$ &$ 0.071 \pm 0.514$\\
LOSS-Union&$573.835/583$ & $0.362 \pm 0.084$& $-0.748 \pm 0.265$ &$ -3.859 \pm 3.982$\\
LU+BAO+CMB&$581.217/592$ & $0.287 \pm 0.009$& $-1.029 \pm 0.118$ &$ -0.109 \pm 0.580$\\
\hline
\multicolumn{5}{c}{FSLLII parametrization}\\
\hline
Constitution&$459.922/394$ & $0.461 \pm 0.032$& $-0.703 \pm 0.350$ &$ -36.447 \pm 21.044$\\
C+BAO+CMB&$468.001/403$ & $0.283 \pm 0.008 $& $-1.008 \pm 0.069$ & $-0.079 \pm 0.461$\\
Union 2&$541.979/554$  & $0.424 \pm 0.061$& $-1.192 \pm 0.188$ &$ -13.110 \pm 16.789$\\
U2+BAO+CMB&$544.517/563$  & $0.282 \pm 0.008$& $-1.058 \pm 0.066$ &$ 0.199 \pm 0.373$\\
Union 2.1&$562.205/577$ & $0.239 \pm 0.406$& $-0.942 \pm 0.648$ &$ 0.531 \pm 3.581$\\
U21+BAO+CMB&$564.349/586$ & $0.282 \pm 0.008$& $-1.033 \pm 0.065$ &$ 0.065 \pm 0.400$\\
LOSS-Union&$574.452/583$ & $0.367 \pm 0.088$& $-1.027 \pm 0.129$ &$ -9.274 \pm 13.042$\\
LU+BAO+CMB&$580.109/592$ & $0.280 \pm 0.008$& $-1.090 \pm 0.055$ &$ 0.371 \pm 0.292$\\
\hline
\multicolumn{5}{c}{SL parametrization}\\
\hline
Constitution&$460.998/394$ & $0.453 \pm 0.0408$& $-0.301 \pm 0.542$ &$ -4.443 \pm 1.930$\\
C+BAO+CMB&$467.949/403$ & $0.284 \pm 0.008 $& $-0.992 \pm 0.098$ & $-1.034 \pm 0.072$\\
Union 2&$541.500/554$  & $0.423 \pm 0.066$& $-0.911 \pm 0.352$ &$ -2.950 \pm 1.933$\\
U2+BAO+CMB&$544.590/563$  & $0.282 \pm 0.008$& $-1.065 \pm 0.087$ &$ -1.014 \pm 0.055$\\
Union 2.1&$562.219/577$ & $0.255 \pm 0.417$& $-0.984 \pm 0.464$ &$ -0.896 \pm 1.574$\\
U21+BAO+CMB&$564.351/586$ & $0.282 \pm 0.008$& $-1.036 \pm 0.086$ &$ -1.019 \pm 0.059$\\
LOSS-Union&$573.840/583$ & $0.353 \pm 0.089$& $-0.806 \pm 0.229 $ &$ -2.198 \pm 1.417$\\
LU+BAO+CMB&$580.632/592$ & $0.281 \pm 0.009$& $-1.099 \pm 0.075$ &$ -1.013 \pm 0.053$\\
\hline
\multicolumn{5}{c}{BA parametrization}\\
\hline
Constitution & 460.793/394 & 0.456$\pm$0.038  & -0.363$\pm$0.508 & -7.596$\pm$4.371 \\
C+BAO+CMB & 467.895/403 & 0.282$\pm$0.008  & -1.006$\pm$0.095 & 0.006$\pm$0.243 \\
Union 2 & 541.576/554 & 0.422$\pm$0.062  & -0.974$\pm$0.301 & -3.460$\pm$3.720 \\
U2+BAO+CMB& 544.570/563 & 0.281$\pm$0.008  & -1.075$\pm$0.088 & 0.163$\pm$0.209 \\
Union 2.1 & 562.187/577 & 0.143$\pm$1.185 & -0.850$\pm$1.294  &  0.407$\pm$1.302 \\
U21+BAO+CMB& 564.274/586 & 0.281$\pm$0.008 & -1.047$\pm$0.086  &  0.103$\pm$0.210 \\
LOSS-Union & 573.933/583 & 0.356$\pm$0.088  & -0.848$\pm$0.206 & -2.442$\pm$2.865 \\
LU+BAO+CMB& 581.294/592 & 0.280$\pm$0.008  & -1.105$\pm$0.075 & 0.215$\pm$0.187 \\
\hline
\end{tabular}
\caption{Best fits for the free parameters using several data sets
for the parametrizations of the EoS of dark energy.}
\label{tab:table01}
\end{table}

To discern which model is the preferred one by observations we only compare the 
$\chi^{2}_{min}$, given in Table \ref{tab:table01}, among data sets.
Notice there is no significant differences among these values for the parametrizations and, 
it is difficult to distinguish which model is the favored by the data. 
Therefore, any parametrization could be a plausible model of dynamical dark energy.

\section{Discussion and Conclusions} \label{sec:conclusions}

In this paper, we have studied five alternative models for dark
energy with an equation of state varying with redshift. We put
constraints on the parameters for these parametrizations using four
different SNIa samples. We reconstructed the cosmological evolution of
the deceleration parameter, and found that using the best fit
obtained with Constitution, Union 2 and LOSS-Union samples, the
Universe reaches a maximum of acceleration at low redshift ($z\sim0.25$),
then it begins to decelerate for all parametrizations like CPL model.
This is certainly a surprise because
although models of the type $w$CDM with $w=$constant, using the same
data, prefers a value $w<-1$ today \cite{Hinshaw:2012, planckXVI, Shafer:2013pxa}, this work
reveals that allowing a variation of $w(z)$ in any of the five
parametrizations studied, implies the deceleration parameter has
already reached its maximum and its evolving towards lower
acceleration regimes.
Nevertheless, this cosmic slowing down of the acceleration
disappears using the constraints derived from the Union 2.1 SNIa
set. This suggests a tension among the different SNIa compilations.
This tension is very intriguing for Union 2 and Union 2.1 because
they differ only in 24 data points and in the lightcurve fitter SALT
for Union 2 and its improved version SALT2 for Union 2.1. Similar tension between 
these Union 2.1 and Union 2 as well as other recent SNIa samples was also found by
\cite{zhang_snia,yang_snia, Shafer:2013pxa}. Additionally, Union 2.1 and LOSS-Union
compilations are very similar and both of them use the lightcurve
fitter SALT2 trained on data from low redshifts SNIa. However, they
predict a reconstructed deceleration parameter considerably
different. Therefore, we conclude the cosmic slowing down of the
acceleration does not depend on the parametrization of the equation
of state of dark energy and it could be due to several factors. 
For instance, systematic errors in the measurements from SNIa
(such as peculiar motions \cite{Hui:2005nm, Gordon:2007zw}), 
the presence of a Hubble bubble, anisotropic cosmological models, to mention some of them
(see \cite{victor_fgas} for a discussion about this point).
As was also discussed in \cite{Cardenas:2014jya}, this low redshift
transition of $q(z)$ was first found by the authors in
\cite{Vanderveld:2006rb}, in the context of a Lemaitre-Tolman-Bondi
inhomogeneous models. In that work, and also in recent ones
(\cite{february2010, Bengochea:2014iha}), the authors derived
an effective deceleration parameter for void models, indicating that
such a behavior of $q(z)$ may be considered as a signature for the
existence of voids.Nevertheless, these results is in contrast with those
obtained when a $q(z)$ parametrization is considered \cite{delCampo:2012ya}.

On the other hand, when adding the BAO and CMB measurements, the
cosmic slowing down of the acceleration disappears for all
parametrizations and they could mimic the cosmological constant at
the present epoch. Therefore, we also confirmed a tension between the
cosmological constraints obtained from low and high redshift data.

In agreement with the results of the recent BAO DR11 measurements
\cite{Delubac:2014aqe} -- where a tension between low to high
redshift observational probes was detected -- and also from the
study in \cite{Cardenas:2014jya} -- where the low redshift
transition of $q(z)$ was demonstrated to be linked with a DE density
that decrease with increasing redshift -- we have also found
evidence for DE density evolution. In fact, using the best fit
parameters found for each parametrization, we can directly plot the
DE density expressions (\ref{eq:xjbp}), (\ref{eq:xfsll}),
(\ref{eq:xsl}) and (\ref{deba}), obtaining for each case the same result found first in 
\cite{Cardenas:2014jya}: the data suggest the DE density is increasing with epoch. 
The astonishing agreement between our approach based on using
low redshift data (as SNIa, $f_{gas}$, and others) with the result
obtained from BAO measurements \cite{Delubac:2014aqe}, make a strong
case for DE evolution, as it was recently highlighted in
\cite{Sahni:2014ooa}.

\begin{acknowledgments}
We thank the anonymous referee for thoughtful suggestions.
We wish to acknowledge useful discussion with Diego Pav\'on and Gael Fo\"{e}x.
J.M. acknowledges support from ESO Comit\'e Mixto, and Gemini $32130024$.
V.M. acknowledges support from FONDECYT $1120741$, and ECOS-CONICYT C12U02.
V.C. acknowledges support from FONDECYT Grant $1110230$ and DIUV 13/2009,

\end{acknowledgments}

\def \prl {Phys.\ Rev.\ Lett. }
\def \ijmpd {Int.\ J.\ Mod.\ Phys.\ D }
\def \prd {Phys.\ Rev.\ D. }
\def \mnras {Mon.\ Non.\ Roy.\ Astron.\ Soc. }
\def \plb {Phys.\ Lett.\ B }

\end{document}